\documentstyle[amssymb,aps,prb]{revtex}

\begin{document}
\author{J.Demsar$^{1}$, R.Hudej$^{1,\ast }$, J.Karpinski$^{2}$, V.V.Kabanov$^{1}$
and D.Mihailovic$^{1}$,}
\address{$^{1}$Institut Jozef Stefan, Jamova 39, 1000 Ljubljana, Slovenia\\
$^{2}$Institut f\"{u}r Festk\"{o}rperphysik, ETH Zurich, Switzerland}
\title{Quasiparticle dynamics and gap structure in Hg$_{1}$Ba$_{2}$Ca$_{2}$Cu$_{3}$O%
$_{8+\delta }$ investigated with femtosecond spectroscopy.}
\maketitle

\begin{abstract}
Measurements of the temperature dependence of the quasiparticle (QP)
dynamics in Hg$_{1}$Ba$_{2}$Ca$_{2}$Cu$_{3}$O$_{8+\delta }$ with femtosecond
time-resolved optical spectroscopy are reported. From the temperature
dependence of the amplitude of the photoinduced reflection, the existence of
two gaps is deduced, one temperature dependent $\Delta _{c}\left( T\right) $
that closes at $T_{c}$, and another temperature independent ''pseudogap'' $%
\Delta ^{p}$. The zero-temperature magnitudes of the two gaps are $\Delta
_{c}\left( 0\right) /k_{B}T_{c}=6\pm 0.5$ and $\Delta ^{p}/k_{B}T_{c}=6.4\pm
0.5$ respectively. The quasiparticle lifetime is found to exhibit a
divergence as $T\rightarrow T_{c}$ from below, which is attributed to the
existence of a superconducting gap which closes at $T_{c}$. Above $T_{c}$
the relaxation time is longer than expected for metallic relaxation, which
is attributed to the presence of the ''pseudogap''. The QP relaxation time
is found to increase significantly at low temperatures. This behavior is
explained assuming that at low temperatures the relaxation of photoexcited
quasiparticles is governed by a bi-particle recombination process.
\end{abstract}


\section{Introduction}

Time-resolved measurements of quasiparticle dynamics in cuprate
superconductors were shown recently to give significant new information
about the single-particle excitations and the low-energy structure of
correlated electron systems such as high-$T_{c}$ superconducting cuprates 
\cite{Kabanov1,ODpaper,Smith} and charge-density wave systems \cite{bb}.
Systematic measurements on YBa$_{2}$Cu$_{3}$O$_{7-\delta }$ (YBCO) have
shown the existence of two simultaneous gaps in the optimally doped and
overdoped region \cite{ODpaper}, but only one (pseudo)gap in the underdoped
region \cite{Kabanov1}. The observation of two energy scales with different
temperature dependences were in apparent agreement with frequency-domain
measurements like angle-resolved photoemission (ARPES) on Bi$_{2}$Sr$_{2}$%
CaCu$_{2}$O$_{8+\delta }$ (Bi2212) \cite{Norman}, as well as Raman
spectroscopy on both YBCO\ and Bi2212 \cite{Hackl}. In this paper we report
the first series of measurements of quasiparticle dynamics on Hg$_{1}$Ba$%
_{2} $Ca$_{2}$Cu$_{3}$O$_{8+\delta }$ (Hg1223) with a $T_{c}$ of 120 K and
find some similarities with femtosecond spectroscopy measurements on YBCO,
but also some differences, particularly the temperature dependence of the
quasiparticle (QP) lifetime at low temperatures.\qquad \qquad

The femtosecond time-resolved pump-probe technique involves the measurement
of small photoinduced changes in the optical reflectivity or transmittance
of a sample caused by photoexcitation. After excitation with a high-energy
photon (1.5 eV), the electrons and holes rapidly relax towards equilibrium;
they scatter amongst themselves and subsequently with lattice phonons in a
process described theoretically by Kaganov et al. \cite{Kaganov} and Allen 
\cite{Allen}. As the carriers reach low energies, the presence of a gap in
the spectrum presents a bottleneck for further relaxation, and the QPs
accumulate at the band edge, waiting to recombine. A second suitably delayed
probe laser pulse measures the change in reflectivity of the sample by
excited state absorption; with these QPs occupying initial states of the
probe transition. Because the QP dynamics critically depends on the presence
of a gap, the technique gives direct information about the temperature
dependence of the QP lifetime and the $T$-dependence of the gap magnitude.
The details of the experimental technique as well as the theory describing
how the gap magnitude is obtained from the data was described in detail
elsewhere \cite{Kabanov1,ACS}.

\section{Experimental details}

The samples used in this investigation were single crystals of Hg$_{1}$Ba$%
_{2}$Ca$_{2}$Cu$_{3}$O$_{8+\delta }$ with a $T_{c}$ of 120K as determined
from AC susceptibility measurements. The samples were prepared in Z\"{u}rich
by a high-gas-pressure synthesis (Ar pressure of 10 - 11 kbar and
crystallization temperatures in the range of 995 $^{o}$C 
\mbox{$<$}%
T 
\mbox{$<$}%
1025 $^{o}$C), with BaCuO$_{2}$-CuO-Ag$_{2}$O eutectic mixture as a flux.
Details of the preparation method are given in Ref. \cite{Karpinski}. As the
result, platelike single crystals of Hg1223 with c-axis thickness of $\sim
10 $ $\mu $m and $0.3\times 0.3$ mm$^{2}$ area were obtained. Great care was
exercised to chose a sample which was without inclusions and was as much as
possible single phase.

The photoinduced reflectivity measurements were performed using a standard
pump-probe technique \cite{ACS}, with a Ti:Sapphire laser producing 70 fs
pulses at approximately 800 nm (1.5 eV) as the source of both pump and probe
optical pulses. The pump and probe pulses were cross-polarized with
polarizations parallel to the $a-b$ plane of the sample. The experiments
were performed at typical pump pulse fluence ${\cal E}_{0}=1.3$ $\mu $J/cm$%
^{2}$ (taking the typical pump pulse energy of 0.1 nJ, and a spot diameter
of $\sim 100$ $\mu $m). The probe intensity was approximately 100 times
lower. Estimating that each absorbed photon with energy $E=1.5$ eV excites $%
N_{QP}=30-40$ quasiparticles ($N_{QP}=E/2\Delta $, where $\Delta $ is the
magnitude of the superconducting gap) and taking the optical absorption
length to be approximately 100 nm, we find the number of photogenerated QPs
due to excitation with a pump pulse to be of the order of $n_{pe}\lesssim
3\times 10^{-3}/$unit cell. On the other hand, the typical carrier
concentration relevant for superconductivity is $n_{0}=2N\left( 0\right)
\Delta \simeq 0.2-0.4/$unit cell, where $N\left( 0\right) $ is the density
of states at $E_{F}$. From the ratio $n_{pe}/n_{0}\lesssim 10^{-2}$ we can
see that we are dealing with weak perturbations of the electronic system and
therefore the pump-probe experiments are probing the equilibrium properties
of the system.

Another important experimental detail that needs to be further discussed is
the sample heating, which takes place due to excitation of the sample with
the train of pump pulses (the heating due to the probe pulse train can be
neglected). In general there are two effects that need to be considered: i)
transient heating due to absorption of the single pulse and ii) steady state
heating, which results in a steady-state temperature increase of the probed
area (in this case the train of pulses separated by 12 ns equals the CW
laser beam with the same average power). It can be shown that in this low
photoexcitation regime the transient heating is on the order of 0.1 K and
can be neglected \cite{JDPhD}. Steady state heating can be quite substantial
($\sim $ 10 K) and should be taken into consideration \cite{ACS,JDPhD} since
the temperature of the probed spot $T_{s}$ may differ substantially from the
temperature of the cold finger $T_{cf}$, which is directly recorded.

The steady state temperature increase, $\Delta T_{cw}=$ $T_{s}-T_{cf}$, can
be accurately determined at temperatures close to T$_{c}$ in several ways 
\cite{ACS,JDPhD}. In the analysis of the data taken on Hg1223 single
crystals we used the so called ''scaling'' method. Namely, in Hg1223 both
amplitude and relaxation time of the photoinduced picosecond reflectivity
transient show anomalies near T$_{c}$ associated with the opening of the
superconducting gap \cite{Kabanov1}. These have thus far already been
observed in various cuprates near optimal doping \cite
{JDPhD,Han,Eesley,Stevens}. When the data obtained with different average
powers of the pump beam are systematically compared, anomalies appear at
different cold finger temperatures. Since the temperature increase is
linearly proportional to the power of the pump beam one can scale the data
and determine the temperature increase near T$_{c}$ quite accurately \cite
{ACS}. With typical experimental conditions $\Delta T_{cw}\left(
T_{c}\right) =9K$ was determined, in close agreement with the calculation
using a heat diffusion model \cite{JDPhD}. Since $\Delta T_{cw}\ $is
inversely proportional to the thermal conductivity $\kappa $ of the sample,
one can estimate the temperature increase in the whole temperature range 
\cite{Carlslaw} providing $\Delta T_{cw}\left( T_{c}\right) $ is known and
the thermal conductivity data is available. In our analysis the thermal
conductivity data from Ref. \cite{Cohn} was used. In all the presented data
temperature increases due to heating of the probed spot were accounted for,
with about $\pm $ 2K uncertainty over the whole temperature range.

\section{Experimental results and data analysis.}

In Fig. 1 we show the time-dependence of the photoinduced signal at a number
of temperatures below and above T$_{c}$. The time-evolution of the
photoinduced reflection $\Delta R/R$ first shows a rapid risetime (of the
order of the pump pulselength) and a subsequent picosecond decay. These data
can be fit quite well using a single exponential decay (see Figure 1) over
the whole temperature range.

In addition to the picosecond decay, we also observe the ubiquitous
near-constant background \cite{KabanovSlow} which has a lifetime longer than
12 ns.

\subsection{Amplitude of the picosecond relaxation component.}

The temperature-dependence of amplitude $\left| \Delta R/R\right| $ and
relaxation time $\tau _{R}$ of the picosecond component in photoinduced
reflectivity are shown in Fig. 2. The picosecond component amplitude, $%
\left| \Delta R/R\right| $, is almost constant at low temperatures, followed
by a rapid decrease as $T_{c}$ is approached. At $T_{c}$ there appears to be
a break in the T-dependence and above $T_{c}$ the amplitude decreases much
more gradually, falling asymptotically to zero at higher temperatures. (At
temperatures above $\sim $ 230 K it is difficult to extract the value of the
amplitude of the picosecond relaxation component, because some very fast
sub-100 fs relaxation becomes evident, which we attribute to metallic
relaxation \cite{Allen,Kaganov,Brorson}.) It is worth mentioning here that
the T-dependence of $\left| \Delta R/R\right| $ [Fig. 2 a)] shows almost
perfect agreement with the data on near-optimally doped YBCO \cite
{Kabanov1,ODpaper,Han,Eesley}.

To analyze the temperature dependence of $\left| \Delta R/R\right| $
quantitatively, we use the model by Kabanov et al. \cite{Kabanov1}, where
the T-dependence of the photoexcited QP density $n_{pe}$ was derived for
different gaps. The model, which was tested also on quasi-1D semiconductor K$%
_{0.3}$MoO$_{3}$ \cite{bb}, is based on the relation 
\begin{equation}
\Delta R=\frac{\partial R}{\partial n}n_{pe}\;.
\end{equation}
The above relation is valid if $n_{pe}$ is small with respect to the number
of thermally excited qp's, which is the case in our experimental
configuration - see Ref.[\cite{Kabanov1,ACS}]. In other words the
perturbation is small and hence linear response is sufficient. Moreover, it
is known that the reflectivity $R$ in HTSC is a very weak function of
temperature in that particular energy range (1.5 eV), as observed
experimentally by Hocomb et al. \cite{Holcomb} using the thermal difference
reflectivity data on variety of high-T$_{c}$ superconductors. Therefore we
expect that the derivative $\partial R/\partial n$ taken in the equilibrium
limit (since we are dealing with linear response) is a weak function of $T$,
and in the first approximation we can consider it to be constant. It should
be pointed out that calculation the constants $\partial R/\partial n$ is the
subject of\ the microscopic theory and probably very model dependent. Our
approach is more phenomenological and requires only that $\partial
R/\partial n$ is weakly temperature dependent.

From the T-dependence of the reflectivity amplitude (which is proportional
to $n_{pe}$) and relaxation time one can, using the analytical expressions
connecting $n_{pe}$ and magnitude and T-dependence of the gap $\Delta (T)$,
extract the magnitude and T-dependence of the gap.

In the limit of small photoexcited carrier density $n_{pe}$, we can assume
that all possible contributions to $\Delta R/R$ - arising from excited state
absorption or photoinduced band-gap changes for example - are linear in the
photoexcited carrier density $n_{pe}$. For an isotropic $T$-dependent gap $%
\Delta _{c}(T),$ the temperature dependence of the amplitude of the
photoinduced reflectivity $|\Delta R/R|$ is given by \cite{Kabanov1}:

\begin{equation}
\left| \Delta R/R\right| \propto n_{pe}=\frac{{\cal E}/(\Delta
_{c}(T)+k_{B}T/2)}{1+\frac{2\nu }{N(0)\hbar \Omega _{c}}\sqrt{\frac{2k_{B}T}{%
\pi \Delta _{c}(T)}}e^{-\Delta _{c}(T)/k_{B}T}}\;,  \label{Eq1}
\end{equation}
where ${\cal E}$ is the incident energy density of the pump pulse per unit
cell, $\nu $ is the effective number of phonon modes interacting with the
QPs, $N(0)$ $\,$is the DOS at $E_{F}$ and $\Omega _{c}\ $is a typical phonon
cutoff frequency. A similar expression gives the amplitude for an isotropic $%
T$-independent gap $\Delta ^{p}$ \cite{Kabanov1}: 
\begin{equation}
\left| \Delta R/R\right| \propto n_{pe}^{\prime }=\frac{{\cal E}/\Delta ^{p}%
}{1+\frac{2\nu }{N(0)\hbar \Omega _{c}}e^{-\Delta _{p}/k_{B}T}}\;.
\label{Eq2}
\end{equation}
In Fig. 2 a) we fit the temperature-dependence of $\left| \Delta R/R\right| $
with the sum of Eq.(\ref{Eq1}) and (\ref{Eq2}) using $\nu =10,$ $\Omega
_{c}=0.1$ eV and $N(0)=5$ eV$^{-1}$cell$^{-1}$spin$^{-1}$. It is evident
from the plots that the total amplitude $\left| \Delta R/R\right| $ cannot
be described by either component separately. However, assuming the
co-existence of two gaps, one of which is $T$-dependent with a BCS-like
temperature dependence and one $T$-independent, we can obtain a good fit to
the data as shown in Fig. 2 a). The values of $\Delta _{c}(0)$ and $\Delta
^{p}$ obtained from the best fit are $6\pm 0.5$ $k_{B}T_{c}$ and $6.4\pm 0.5$
$k_{B}T_{c}$\ respectively.

At this point we should state that a simple prediction for the case of $d$%
-wave gap (gapless DOS) cannot account for the observed data. Namely, as
soon as we assume that density of states is gapless, then after the initial
relaxation QPs would accumulate in the nodal regions and the number of
photoexcited QP's can be approximated as $n_{pe}$ $=E_{i}/T^{\ast }$, where $%
T^{\ast }$ is their effective temperature. This means that sub-linear
dependence of the photoinduced signal amplitude as a function of
photoexcitation intensity $E_{i}$ should be observed in case of gapless
density of states. This is due to the temperature dependence of the
electronic specific heat in case of a d-wave superconductor, which goes as $%
C_{el}\propto T^{2}$. This clearly contradicts our experiments, since linear
dependence of photoinduced reflectivity amplitude on $E_{i}$ was observed
over wide range of photoexcitation energies - see Ref. \cite{Kabanov1,linear}%
. Moreover, the d-wave model (see Ref. \cite{Kabanov1} for details)
predicts\ non-exponential relaxation, no-anomaly at $T_{c}$ in the
relaxation time and different T-dependence of the picosecond component
amplitude, all of which are not consistent with the data.

\subsection{Relaxation time.}

The T-dependence of $\tau _{R}$, obtained by the single exponential decay
fit to the data is shown in Fig. 2 b). There are two noticeable features in
the observed temperature dependence. First, at temperatures above $\sim $ 80
\ K the T-dependence is quite similar to the behavior seen in optimally
doped and overdoped YBCO \cite{Kabanov1,ODpaper,Han,Eesley}. Upon increasing
temperature through $T_{c}$ the relaxation time [see inset to Fig. 2 b)]
shows an anomaly. Such an anomaly is expected to occur in the relaxation
time in the presence of a gap which closes at $T_{c}$ \cite{Kabanov1}. Near T%
$_{c}$ the relaxation is governed by the anharmonic decay time of high
frequency phonons \cite{Kabanov1} given by: 
\begin{equation}
\tau _{ph}=\frac{\hbar \omega ^{2}\ln \left\{ 1/({\cal E}/2N(0)\Delta
_{c}(0)^{2}+e^{-\Delta _{c}(T)/k_{B}T})\right\} }{12\Gamma _{\omega }\Delta
_{c}(T)^{2}}\;,  \label{Tph}
\end{equation}
$\Delta _{c}\left( T\right) $ being the magnitude of the T-dependent gap, $%
\omega $ is the phonon frequency (typically $\omega \simeq 500\ $cm$^{-1}$),
and $\Gamma _{\omega }$ is the optical phonon linewidth, typically $10$ cm$%
^{-1}$. Near T$_{c}$ $\tau _{ph}\propto 1/\Delta _{c}\left( T\right) $ as
plotted in inset to Fig. 2 b). Divergence below $T_{c}$ is followed by a
rapid drop to a lower value, and above T$_{c}$ the relaxation time shows
only a weak T-dependence. It needs to be mentioned that above $T_{c}$, $\tau
_{R}$ is much longer than expected for metallic relaxation \cite
{Kabanov1,Brorson}, implying the presence of a pseudogap in the density of
states as already observed in YBCO \cite{ODpaper} and consistent with the
observed T-dependence of $\left| \Delta R/R\right| $. At temperatures above $%
\sim $ 230 K an additional very fast relaxation with sub-100 fs relaxation
time becomes evident, which we attributed to metallic relaxation. At
temperatures below $\sim $ 80 K, unlike in YBCO \cite{Kabanov1}, $\tau _{R}$
shows a strong T-dependence $-$ increasing rapidly as temperature is
decreased. A similar T-dependence has been reported also in Bi2212 \cite{Gay}
and Tl$_{2}$Ba$_{2}$CuO$_{6+\delta }$ (Tl2201) \cite{Smith}. The possible
origin of different low temperature behavior in various HTSC will be
discussed in the Discussion.

\subsection{The amplitude of nanosecond component.}

In Fig. 3 we show the T-dependence of the long-lived photoinduced signal
amplitude ${\cal D}$ (see Fig. 1). The lifetime appears to be much longer
than the distance between two successive pump laser pulses, so its
relaxation time cannot be measured directly. However, from the comparison of
the amplitude at negative time delays ($\sim 12$ ns after photoexcitation)
with the photoinduced signal at 100 ps (when there is no picosecond
relaxation signal left) one can estimate the relaxation time to be on the
order of 100 ns or longer \cite{KabanovSlow}. Indeed, similar photoinduced
signal was observed also on 100 $\mu s$ timescale \cite{Thomas}, which is
close to the measured $\mu s$ dynamics in the photoinduced absorption of
mm-waves \cite{Feenstra}. Taken all together it seems we are dealing with
glass-like relaxation dynamics with no well defined timescale \cite{Phillips}%
.

At low temperatures the signal amplitude increases upon increasing
temperature which is contrary to the expected behavior due to heating \cite
{JDPhD}. At T$_{c}$ the amplitude drops substantially followed by a gradual
decrease at higher temperatures. The nanosecond relaxation component was
previously observed also on YBCO \cite{Stevens} and the quasi 1D CDW
semiconductor K$_{0.3}$MoO$_{3}$ \cite{bb}. Its lifetime and T-dependence
suggested an explanation for the long lived signal in terms of in-gap
localized states \cite{KabanovSlow}. The model gives the T-dependence of
slow component amplitude for different T-dependences of the gap. In case the
gap is mean-field like, closing at $T_{c}$ the T-dependence of ${\cal D}$ is
given by \cite{KabanovSlow} 
\begin{equation}
{\cal D}\propto \sqrt{\eta n_{pe}(T)}\frac{\Omega _{c}}{\Delta _{c}(T)}\;,
\label{Eq3}
\end{equation}
where $\eta =\gamma \tau _{ph}$ $\propto $ $\eta ^{\prime }/\Delta
_{c}\left( T\right) $ is the probability of trapping a QP into a localized
state, $n_{pe}(T)$ is the number of photoinduced QPs at temperature $T$
created by each laser pulse given by Eq.(\ref{Eq1}), $\Omega _{c}$ is the
phonon cutoff energy and $\Delta _{c}(T)$ is the T-dependent gap magnitude.
In case the gap $\Delta ^{p}$ is T-independent, the model gives \cite
{KabanovSlow} 
\begin{equation}
{\cal D}\propto \sqrt{\frac{\eta n_{pe}^{\prime }(T)}{\alpha
(1-(T/T_{c}))+\Gamma }},\Biggm\{%
{T<T_{c}\,;\,\alpha >0 \atop T>T_{c}\,;\,\alpha =0}%
\label{Eq4}
\end{equation}
Here $n_{pe}^{^{\prime }}$ is given by Eq.(\ref{Eq2}) and $\Gamma $ is the
T-independent bi-particle recombination time present also at temperatures
above T$_{c}$ since $\Delta ^{p}$ is T-independent. Below $T_{c}$ the
presence of the condensate may also have an effect on the recombination of
localized excitations, which gives first term proportional to square of the
order parameter in the denominator \cite{KabanovSlow}.

The data [see Fig. 3] show substantial long-lived photoinduced signal
present also at temperatures above $T_{c}$, suggesting the presence of gap
(and in-gap localized states) also at high temperatures, similar as deduced
from the picosecond relaxation data - see Fig. 2. We thus fit the data with
the model \cite{KabanovSlow} assuming the co-existence of two gaps, a
T-dependent gap $\Delta _{c}\left( T\right) $ and a T-independent
(pseudo-)gap $\Delta ^{p}$, with the two contributions to the signal ${\cal D%
}$ given by Eqs.(\ref{Eq3}) and (\ref{Eq4}) respectively. Substituting $%
\Delta _{c}\left( 0\right) $ and $\Delta ^{p}$ from fits to the picosecond
decay components into Eqs.(\ref{Eq3}) and (\ref{Eq4}) we obtain the solid
line fit in Fig. 3. The two components are plotted separately by dashed [Eq.(%
\ref{Eq3})] and dotted [Eq.(\ref{Eq4})] lines respectively.

\section{Discussion}

The present measurements appear to show very similar 2-gap behavior as in
overdoped and optimally doped YBCO \cite{ODpaper}. In particular, both show
an apparent co-existence of two gaps for QP excitations, that is, a
pseudogap $\Delta ^{p}$ coexisting with a collective temperature-dependent
gap $\Delta _{c}(T)$ which closes at $T_{c}$. Unlike in YBCO, where the two
relaxation components (one present also above $T_{c}$ with T-independent $%
\tau _{1}$, and the other present only at $T<T_{c}$ with $\tau _{2}$
diverging at $T_{c}$\cite{ODpaper}) are clearly distinguishable in the decay
because of their very different lifetimes, on Hg1223 the relaxation is well
reproduced by single exponential decay. However the presence of picosecond
timescale relaxation above $T_{c}$, together with an asymptotic decrease in $%
\left| \Delta R/R\right| $ at high temperatures, suggests similar
two-component behavior, with the two relaxation times too close to be
resolved. This is supported by the fact that the relaxation time at $T>T_{c}$
in Hg1223 is nearly the same as at $T<T_{c}$ ($\sim 100$ K), whereas in YBCO
it is found to be almost an order of magnitude lower \cite{ODpaper}. Two
such distinct picosecond relaxation components with opposite signs were
observed also on Tl$_{2}$Ba$_{2}$Ca$_{2}$Cu$_{3}$O$_{10}$ (Tl2223) \cite
{Eesley}, BI2212 \cite{Gay} and Tl2201 \cite{Smith}, suggesting that the two
component behavior is quite general in high-$T_{c}$ superconductors near
optimal doping. Furthermore, the two gap behavior is clearly apparent in the
slow component T-dependence as discussed in the previous section. We note
that the apparent similarity in all these HTSC materials is very important
from the theoretical point of view of universality of the low-energy
excitations in cuprate superconductors. More specifically, the apparently
universal coexistence of two components (two gaps) in YBCO, BI2212, Hg1223,
Tl2201 and Tl2223 appears to impose some quite stringent restrictions on the
theoretical framework for the solution of the high-$T_{c}$ problem.

The main difference between the data on Hg1223 and optimally doped YBCO is
in the behavior of the relaxation time at low temperatures. While in YBCO $%
\tau _{R}$ is almost T-independent at low temperatures \cite{Han,Kabanov1},
showing only slight upturn at very low temperatures \cite{Han,ODpaper}, in
Hg1223 $\tau _{R}$ increases significantly upon lowering the temperature
from $\sim 80$ K. A similar T-dependence was observed also in Tl2201 \cite
{Smith} and Bi2212 \cite{Gay}, suggesting that this low-T behavior of $\tau
_{R}$ is not universal.

In order to understand the low temperature behavior of $\tau _{R}$ in
Hg1223, and account for the difference in behavior between YBCO and Hg1223,
we have analyzed the processes governing the relaxation of photoexcited
quasiparticles in more detail.

We first consider the theoretical model for quasiparticle relaxation \cite
{Kabanov1}, which quantitatively described the T-dependence of QP relaxation
in YBCO \cite{Kabanov1} as well as in quasi 1D CDW semiconductor K$_{0.3}$MoO%
$_{3}$ \cite{bb} over wide range of temperatures. The model assumes that
after excitation with a high-energy photon, the electrons and holes rapidly
relax towards equilibrium; they scatter amongst themselves (quasiparticle
avalanche multiplication due to electron-electron collisions) and
subsequently with lattice phonons reaching states just above the band edge
within $\tau \ll 100$ fs \cite{Allen,Kaganov,Brorson}. The photoexcited QPs
recombine with the creation of high frequency phonons with $\hbar \omega
>2\Delta $. High frequency phonons, on the other hand, get reabsorbed
creating new pairs of QPs, or anharmonicaly decay into low energy phonons ( $%
\hbar \omega <2\Delta $) which cannot excite new QPs because of energy
conservation. In case the recombination and reabsorption processes are fast
compared to the anharmonic phonon decay (typically a few ps) a near-steady
state distribution of the QPs and{\em \ }high frequency phonons form is
established on a sub-$100$ fs timescale, described by common temperature.
The relaxation rate of the photoinduced QPs is then dominated by the energy
transfer from high-frequency phonons with $\hbar \omega >2\Delta $ to
phonons with $\hbar \omega <2\Delta $ \cite{Kabanov1} given by Eq.(\ref{Tph}%
).

The assumption that recombination is fast compared to the anharmonic phonon
decay, however, can be violated at low temperatures, when the gap is large
and the number of thermally excited QPs is small. It can lead to a situation
when the recombination time becomes longer than the anharmonic phonon decay
time. In this case the relaxation time of photoexcited QP density is
governed by bi-particle recombination process, and QPs and high energy
phonons can be described by quasi-equilibrium distribution functions with
different temperatures $T_{qp}$ and $T_{ph}$. (Note that both temperatures
are higher than the equilibrium lattice temperature $T$, which is also the
temperature of the low energy phonons \cite{Kabanov1}.)

To estimate the temperature dependence of the relaxation time of
photoexcited QP density (governed by bi-particle recombination process) at
low temperatures, we consider the kinetic equation for QPs \cite{Aronov}.
The collision integral describing the kinetics of the QPs has two different
terms. The first one describes inelastic scattering of QPs (with creation or
absorption of a phonon) and the second describes the recombination (or
creation) of two QPs with creation (or absorption) of a high-frequency
phonon $(\hbar \omega >2\Delta )$. The ratio of these two terms is
determined by coherence factors, and when $T\ll T_{c}$, the first term is
small as $\sim \left( T/\Delta \right) ^{2}$. In this case the rate equation
for the total density of QPs, $n(T_{qp})$, can be reduced to the equation
(see also Ref. \cite{kresin}): 
\begin{equation}
\frac{\partial n(T_{qp})}{\partial t}=\frac{8\pi \lambda \Delta ^{2}}{\hbar
\Omega _{c}N(0)}[n^{2}(T_{qp})-n^{2}(T_{ph})]\;.  \label{V1}
\end{equation}
As can be seen from this equation when $T_{qp}\gg T_{ph}$ the relaxation is
non-exponential. On the other hand, when $T_{qp}\approx T_{ph}$, we can
linearize the right hand side of Eq.(\ref{V1}) to obtain: 
\begin{equation}
\frac{\partial n(T_{qp})}{\partial t}=\frac{1}{\tau _{rec}}%
[n(T_{qp})-n(T_{ph})]\;,
\end{equation}
with 
\begin{equation}
\tau _{rec}=\frac{\hbar \Omega _{c}^{2}}{32\pi \lambda \Delta ^{2}\sqrt{\pi
\Delta k_{B}T_{ph}/2}}\exp {(\Delta /k}_{B}{T_{ph})\;.}  \label{Vfin}
\end{equation}
Here $\lambda $ is the dimensionless electron-phonon coupling constant, for
HTSC typically on the order of 1 \cite{Brorson}. As a result, $\tau _{rec}$
shows an exponential increase at low temperatures, whereas the relaxation
time of the high-frequency phonons is constant at low $T$ \cite{Kabanov1}.
It means that at some temperature $T<T_{c}$ we should expect a crossover
from high temperature relaxation behavior near $T_{c}$, as described
previously in Ref. \cite{Kabanov1}, to low temperature recombination which
shows different behavior and is described by Eq.(\ref{Vfin}).

In Fig. 4 we plot the relaxation time data, together with expressions (\ref
{Tph}) and (\ref{Vfin}) describing the T-dependence of the relaxation times $%
\tau _{ph}$ and $\tau _{rec}$ respectively. At temperatures close to $T_{c}$
the divergence in relaxation time $\tau _{R}$ is well reproduced by $\tau
_{ph}$, whereas at temperatures below $\sim 70$ K $\tau _{R}$ becomes larger
than predicted $\tau _{ph}$.

To be able to compare the low temperature relaxation time data on Hg1223
with Eq.(\ref{Vfin}), there is an important detail that needs further
discussion. Namely, the temperatures $T_{qp}$ and ${T_{ph}}$ entering Eqs.(%
\ref{V1}) and (\ref{Vfin}) depend on $T$ and $\Delta $ \cite{Kabanov1}. Near 
$T_{c}$, when the gap is small, the number of photoexcited carriers is small
compared to the number of thermally excited QPs (the same goes also for
densities of high energy phonons) therefore $T_{qp}$ and ${T_{ph}}$ are very
close to the equilibrium lattice temperature $T$. At low temperatures, on
the other hand, the situation is changed and even weak photoexcitation
strongly increases $T_{qp}$ (and ${T_{ph}}$) with respect to the equilibrium
temperature $T$. Assuming that all the absorbed energy goes to quasiparticle
system one obtains \cite{Kabanov1} 
\begin{equation}
k_{B}T_{qp}\simeq \Delta (T)/\ln \left\{ 1/({\cal E}/2N(0)\Delta
^{2}(0)+\exp (-\Delta (T)/k_{B}T))\right\}  \label{Tqp}
\end{equation}
giving $T_{qp}\simeq T_{c}/2$ for the limiting case when $T\rightarrow 0$
using the above experimental configuration \cite{Kabanov1}. In case the
anharmonic phonon decay is faster than bi-particle recombination time ${%
T_{ph}}$ is expected to be lower than $T_{qp}$, expression \ref{Tqp} giving
an upper limit for $T_{ph}$. Since the main T-dependence of $\tau _{rec}$
[Eq.(\ref{Vfin})] comes from the $\exp {(\Delta /k}_{B}{T_{ph})}$ small
changes in ${T_{ph}}$ bring substantial change in $\tau _{rec}$, therefore
this surely is an important issue. In Fig. 4 we plot Eq.(\ref{Vfin}) using
two extreme cases: dotted line represents expression (\ref{Vfin}) with $%
T_{ph}=T$, whereas solid line represents Eq.(\ref{Vfin}) where $T_{ph}$ is
given by Eq.(\ref{Tqp}) and plotted in inset to Figure 4. As can be seen
both fits reasonably well account for the data, giving $\Delta
/k_{B}T_{c}\approx 2-4$ depending strongly on the ${T_{ph}(T)}$. At
temperatures below $\sim 30$ K $\tau _{rec}$ is expected to saturate.
However at low temperature $T_{qp}$ is expected to be substantially higher
than $T_{ph}$ leading to non-exponential relaxation. Indeed the crossover to
non-exponential relaxation was reported at very low temperatures in Bi2212 
\cite{Gay} and Tl2201 \cite{Smith}. Considering that the T-dependence of $%
\tau _{rec}$ is governed by $\exp {(\Delta /k}_{B}{T_{ph})}$, a crossover
from high temperature {\it relaxation} to low temperature {\it recombination}
picture is expected to highly depend on the magnitude of superconducting gap 
$\Delta $. Since the gap value in YBCO, determined from tunneling data is
lower than that of Bi2212 (and Hg1223) the crossover is expected to be lower
in temperature.

\section{Conclusions}

We have performed measurements of the temperature dependence of the
quasiparticle dynamics in Hg$_{1}$Ba$_{2}$Ca$_{2}$Cu$_{3}$O$_{8+\delta }$
with femtosecond time-resolved optical spectroscopy. From the temperature
dependence of the amplitude and relaxation time of the photoinduced
reflection, the existence of two gaps is deduced, one temperature dependent $%
\Delta _{s}\left( T\right) $ that closes at $T_{c}$, and a temperature
independent pseudogap $\Delta ^{p}$. The zero-temperature magnitudes of the
two gaps obtained from the fit to the data using theoretical model by
Kabanov et al. \cite{Kabanov1} are $\Delta _{s}\left( 0\right)
/k_{B}T_{c}=6\pm 0.5$ and $\Delta ^{p}/k_{B}T_{c}=6.4\pm 0.5$ respectively.
In addition to picosecond quasiparticle relaxation component a long lived
nanosecond component was observed, whose dynamics is described with the
model for photoexcited localized in-gap state relaxation \cite{KabanovSlow}.

Unlike in YBCO \cite{ODpaper} the relaxation of the picosecond transient is
found to be single exponential over the whole T region, showing significant
increase at low temperatures. From the model analysis we suggest that at low
temperatures the relaxation time is dominated by bi-particle recombination
in this material and find good agreement between the model and the data. The
fact that the relaxation times for the pseudogap relaxation and collective
gap relaxation are nearly the same in Hg1223, while in YBCO they differ by
almost an order of magnitude still needs to be understood.

\newpage

\section{Figure captions}

Figure1:

a) $\Delta $R/R taken on Hg-1223 at 70 K. The fast risetime is followed by a
picosecond decay. Some long-lived photoinduced signal (difference between
signal at point ${\cal D}$ and zero signal, when pump beam is blocked)
persists up to 12 ns (difference between two successive pump pulses),
resulting in a temperature-dependent offset. b) $\Delta $R/R at various
temperatures below and above T$_{c}$. In these traces the T-dependent
background, ${\cal D}$, is subtracted. (Its T-dependence is analyzed
separately.) Inset: Real part of the AC susceptibility taken on one of the
samples, showing a T$_{c}$ onset at 120 K.

Figure 2:

a) Amplitude of the picosecond component, $\left| \Delta R/R\right| $, fit
with the sum of two components (solid line) given by Eq.(1) - dashed, and
Eq.(2) - dotted. Values of $\Delta _{c}\left( 0\right) $ and $\Delta ^{p}$
from the fit are shown. b) The temperature dependence of relaxation time $%
\tau _{R}$. Inset: the divergence at $T_{c}$ is compared to the $1/\Delta
_{c}\left( T\right) $, with $\Delta _{c}\left( T\right) $ having BCS
T-dependence.

Figure 3:

The T-dependence of the slow component amplitude ${\cal D}$ [see Fig 1 a)],
together with the fit using theoretical model for photoinduced absorption
from in-gap localized states - see text.

Figure 4:

The relaxation time data compared to the theoretical predictions. Eq.(\ref
{Tph}) is plotted by dashed curve, whereas expression (\ref{Vfin}) with $%
T_{ph}=T$ and $T_{ph}\ $given by Eq.(\ref{Tqp}) is plotted with dotted and
solid line respectively. At temperatures below $\sim 30K$ Eq.(\ref{Vfin}) is
expected to fail, and the relaxation becomes non-exponential. Inset: The
T-dependence of the high-energy phonon temperature ${T_{ph}}$ for $T_{ph}=T$
(dotted) and $T_{ph}$ given by Eq.(\ref{Tqp}) (solid).

\newpage 
\footnotetext{
* current address: Polytechnik, Nova Gorica, Slovenia}


\begin{references}
\bibitem{linear}  In our measurement we find $\left| \Delta R/R\right| $ to
be linearly dependent on pump pulse fluence ${\cal E}$ over an order of
magnitude in ${\cal E}$ .
\end{references}
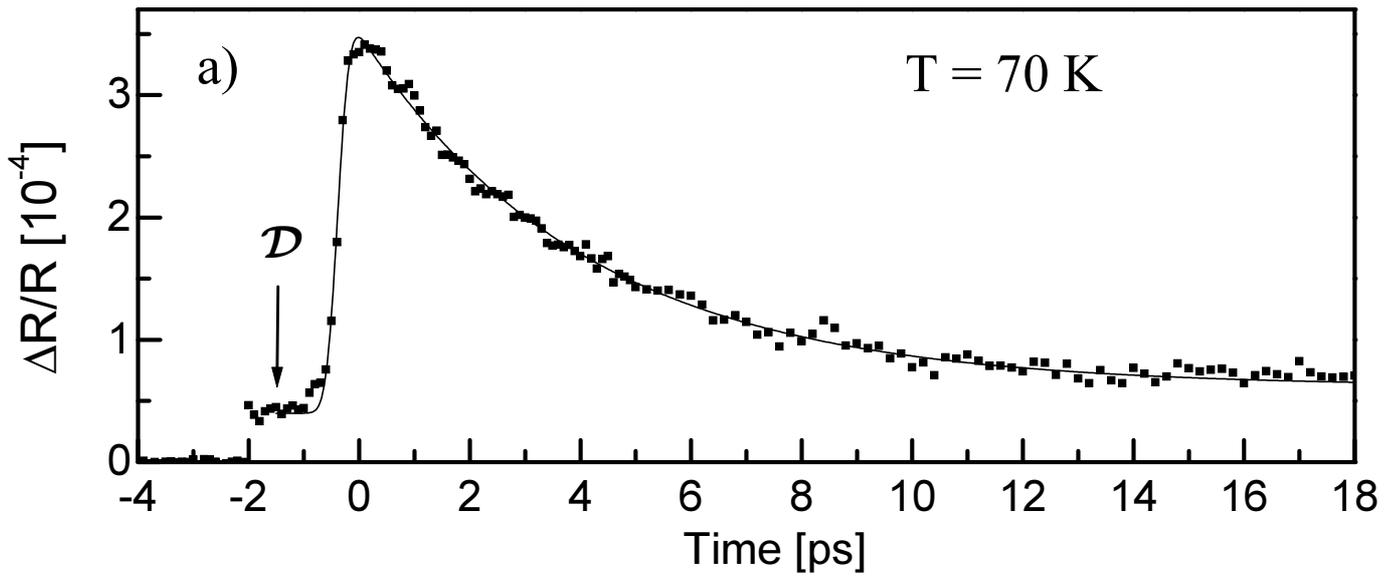
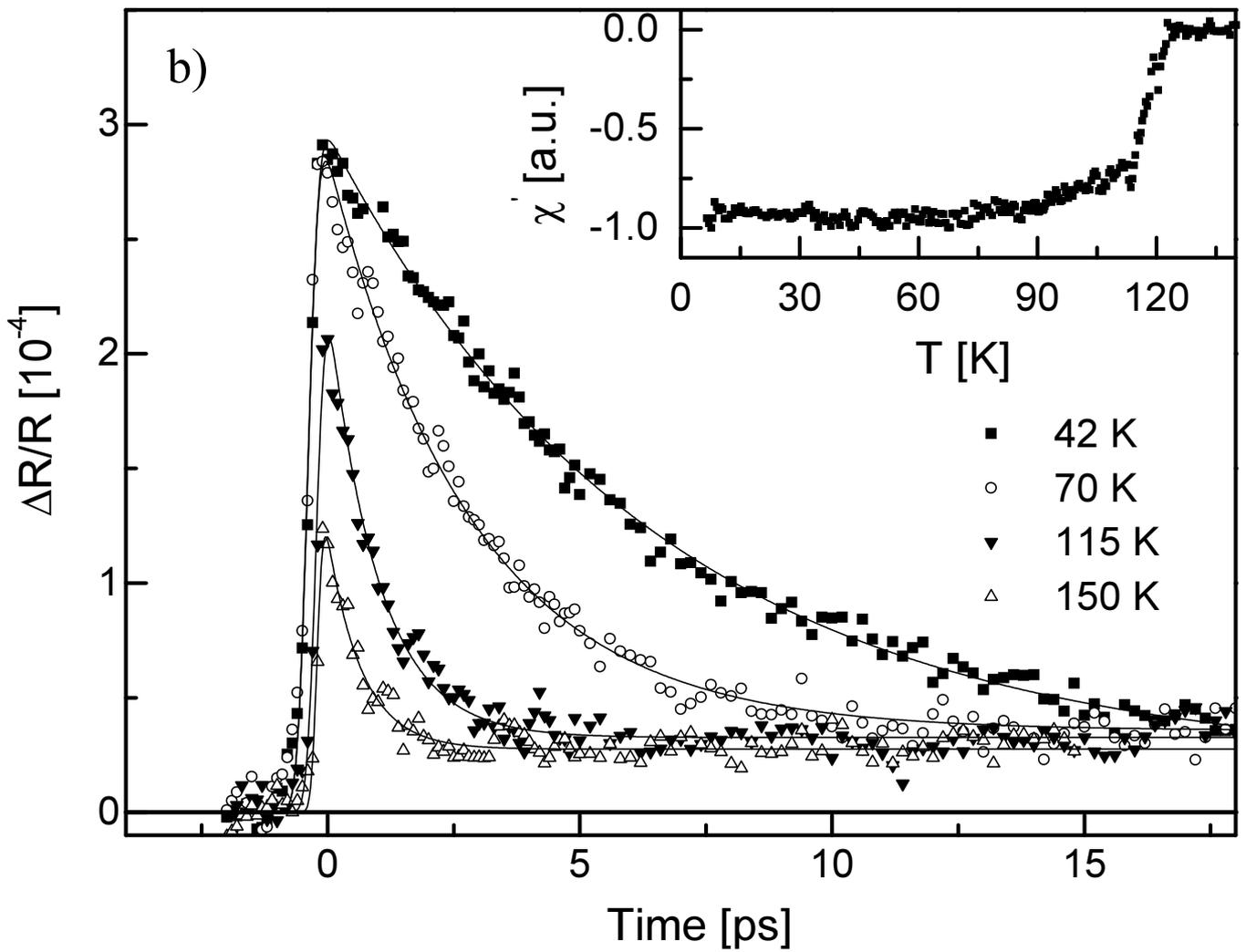
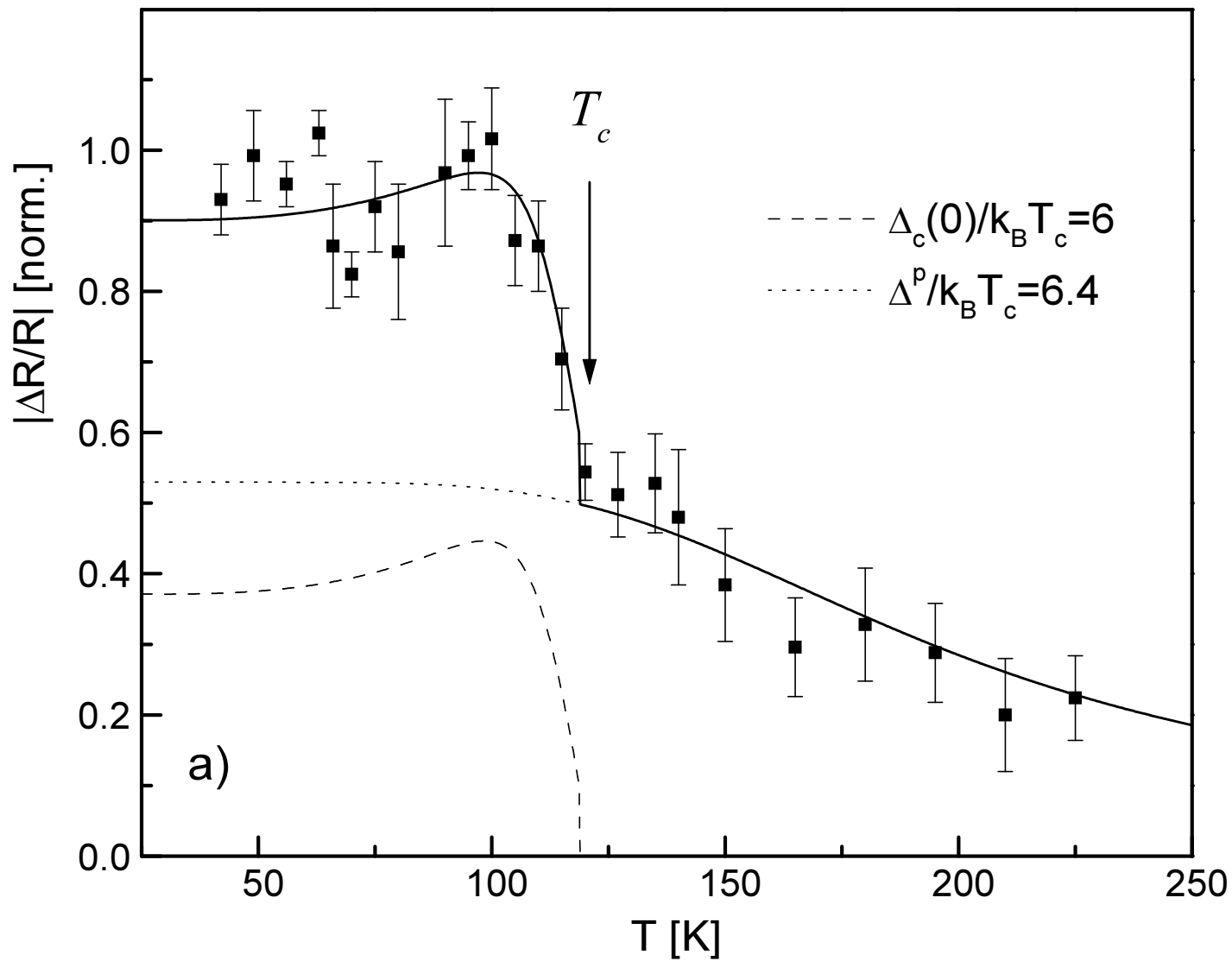
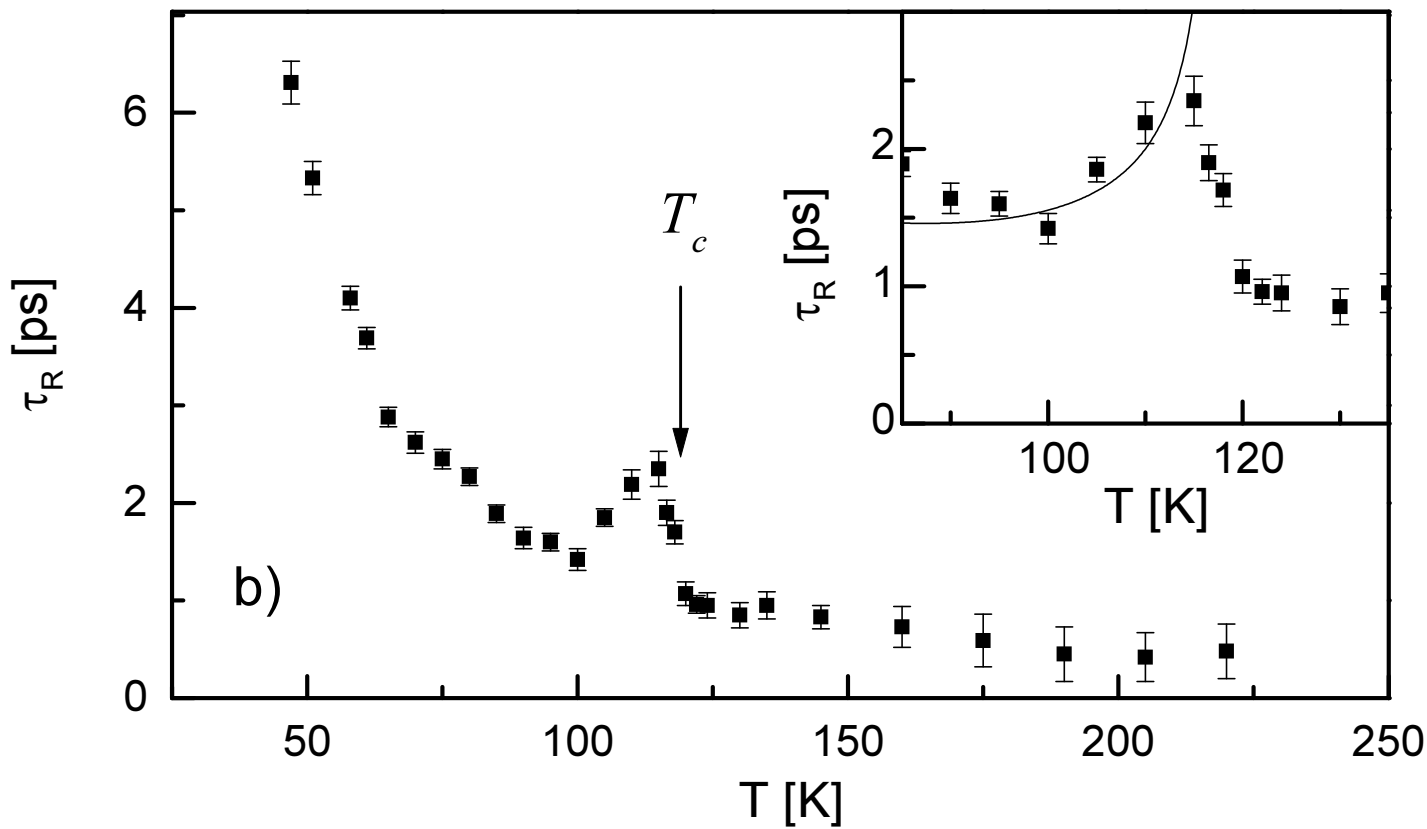
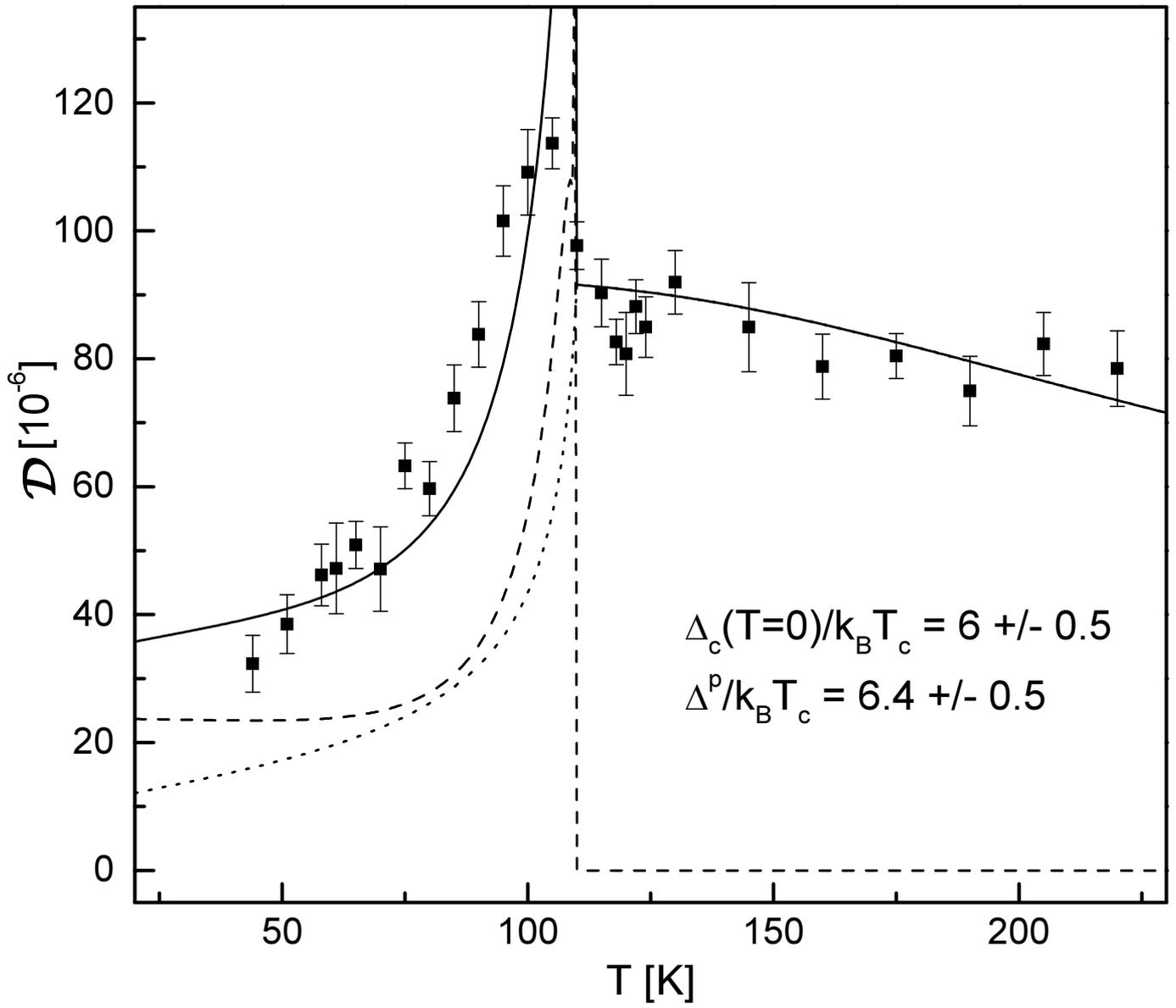
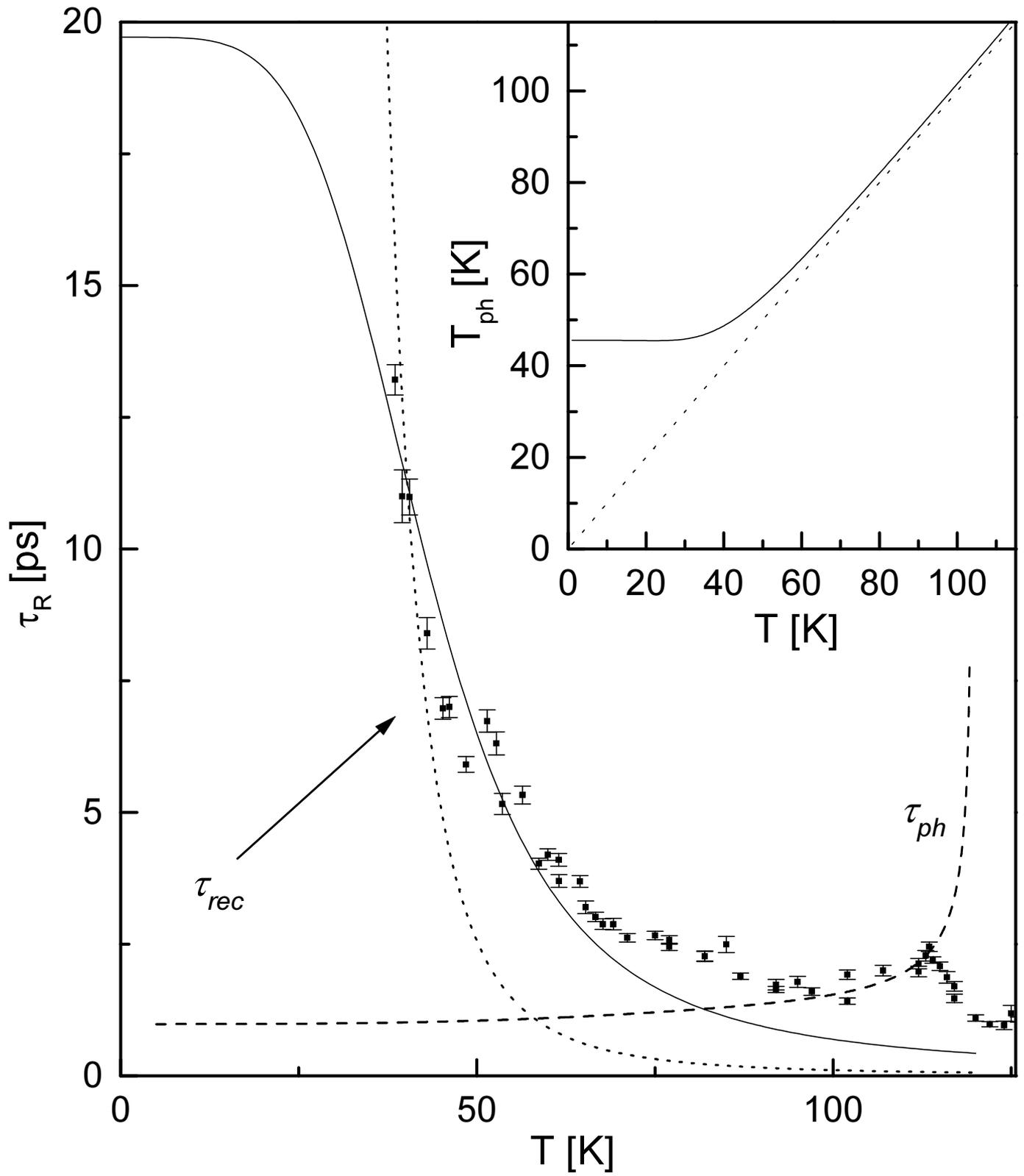
\end{document}